# Performance advantages of buffered mode operation of HxRG near infrared detectors

Naidu Bezawada, Derek Ives, Domingo Alvarez, Benoit Serra, Elizabeth George, Christopher Mandla and Leander Mehrgan

European Southern Observatory, Karl-Schwarzschild-Strasse 2, 85748 Garching bei München, Germany

## ABSTRACT

The Teledyne HxRG detectors have versatile and programmable output options to allow operation of them in a variety of configurations such as slow unbuffered, slow buffered, fast buffered or unbuffered modes to optimise the detector performance for a given application. Normally at ESO, for low noise operation, the detectors are operated in slow unbuffered mode. Whilst the slow unbuffered mode offers a simple interface to the external preamplifier electronics, the detector operation in this mode can suffer from reduced pixel frequency response and higher electrical crosstalk between the readout channels. In the context of the detector systems required for the first generation instruments of the ELT (MICADO, HARMONI and METIS), an exercise was undertaken to evaluate the noise, speed and crosstalk performance of the detectors in the slow buffered mode. A test preamplifier has been designed with options to operate a H2RG detector in buffered or unbuffered and with or without using the reference output, so a direct performance comparison can be made between different modes. This paper presents the performance advantages such as increased pixel frequency response, elimination of electrical crosstalk between the readout channels and the noise performance in the buffered mode operation. These improvements allow us to achieve the same frame readout time using half the detector cryogenic electronics and detector controller electronics for the ELT instruments, which significantly reduces the associated cryo-mechanical complexities in the instrument.

**Keywords:** Infrared detectors, H2RG, H4RG, Buffered Output Mode Operation, Differential preamplifier, Performance of near infrared detectors

## 1. INTRODUCTION

The Hawaii1/2/4RG detectors can be operated in different output modes such as slow un-buffered mode, slow buffered mode and fast buffered or unbuffered mode by programming appropriate registers on the device. For low noise, the detectors are operated in slow un-buffered mode, where the detector pixel unit cell is directly connected to the external cryogenic preamplifiers located close to the detector. As the pixel unit cell source follower has limited signal drive capability, it is necessary to have the preamplifiers close to the detector to reduce the external load capacitance. The detectors in the slow un-buffered mode are operated at about 100kHz or less pixel rate. Whilst the un-buffered mode provides a simple interface to the external preamplifier electronics, the pixel frequency response becomes poor and electrical crosstalk between the readout channels increases to unacceptable levels at higher pixel frequencies due to high output impedances. These limitations can be overcome when the detector is operated in slow buffered mode at the expense of a small increased detector power dissipation. The output source follower buffers provide low output impedance and isolate the external load capacitance from the pixel unit cell, there by achieving faster pixel settling time, hence higher pixel frequency response and reduced signal coupling between the readout channels.

A test preamplifier has been designed to test the detector in different output modes in slow speed in order to compare the performance of the detector. This paper presents the performance advantages of a H2RG detector in slow buffered mode operation over slow unbuffered mode of operation. Section 2 presents the details of the buffered mode operation and the differential pre-amplifier configuration of the test preamplifier. The section also presents a new CMOS operational amplifiers which have many advantages over the current CMOS amplifiers which are widely used at ESO and many other observatories. Section 3 presents the results from a warm multiplexer that already show these performance advantages while section 4 presents results from a cold testing of a science H2RG detector. Section 5 presents a simple model for reduced electrical crosstalk between the readout channels.

## 2. BUFFERED MODE OPERATION

Figure 1 below shows a block diagram of the pixel unit cell / column bus and a configurable detector output stage for the slow mode. When the detector is operated in slow un-buffered mode, the source follower (SF) in the pixel unit cell directly drives the signal though the column bus to the detector output pads. The output impedance of the detector in this mode is high and is typically a few kilo ohms at 70K to a few tens of kilo ohms at room temperature. In the buffered operation, the on-chip PMOS buffers isolate the pixel source follower from the external load capacitance and provide low output impedance outputs, to only a few hundred ohms at 70K. However, in the buffered operation each output needs to be provided with suitable amplifier load current for their operation, which is provided by the cryogenic preamplifier board (as shown in the dotted block). This was successfully verified with H1RG detector[1] albeit following the publication of the results.

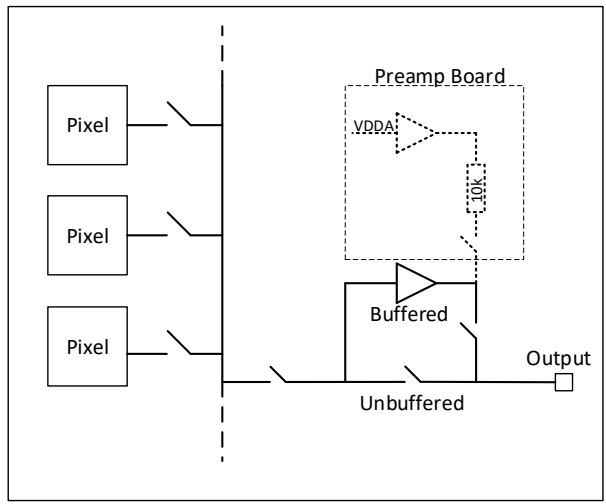

Figure 1. Block diagram of pixel and output configuration.

### 2.1 Test preamplifier

A test preamplifier (Figure 2) has been designed to test a H2RG detector for its speed, noise and crosstalk performance in the buffered mode and to make comparisons with the conventional unbuffered mode of operation. The preamplifier is designed with only 12 differential amplifier channels to verify the concepts. Output channels are chosen such that it allowed to read out one quarter of the H2RG detector using readout channels 0 to 7, as well as a full frame using readout channels 7, 15, 23 and 31 output channels. A readout through output channels 0 to 7 allow to measure any electrical crosstalk between the consecutive readout channels. It includes options to operate the detector in slow un-buffered or buffered mode via a selection of jumpers on the board. In addition, the reference output is also digitised in parallel with the normal output channels. In the buffered mode the outputs are pulled to individual pull-up supplies for each output through 10k resistors providing approximately 100uA drain current through the PMOS output buffers. The detector analog supply is individually buffered and supplied with a pull-up resistor for each output. A differential gain of 8.10 has been implemented in the preamplifier. The 3-dB bandwidth of the preamplifier is set at about 720kHz using RC filter across the feedback of the op-amps.

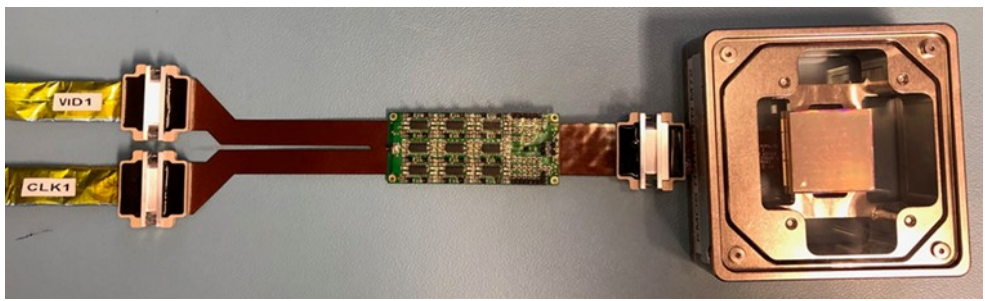

Figure 2. Test preamplifier connected to a H2RG multiplexer.

**2.2 New cryogenic CMOS op-amps**

The standard ESO preamplifiers used over two decades are based on Texas Instruments CMOS amplifiers, TLC2274 for slow speed operation up to 100kHz with the required gain. These amplifiers have a gain bandwidth product (GBW) of about 2.6MHz. High speed amplifiers such as OPA4350 or OPA4354 are used for high speed operation. These high speed amplifiers have 35 MHz / 100 MHz GBW and consume higher power and are not ideal for slow speed operation. For the test preamplifier, a more suitable, low noise, low power, low input offset and ultralow drift quad CMOS op op-amps (OPA4192) have been selected to test the detector in slow speed up to 500kHz pixel rates.

The OPA4192 has a bandwidth of 10MHz with a sufficient slew rate (20V/µs) such that it meets the speed and gain requirements for operation up to 500kHz pixel rate in slow speed. The low input offset and drift offer an excellent DC performance whilst its low noise and high capacitive drive capability makes it attractive for cryogenic instruments where output signals are driven over long cable lengths. The OPA4192 is based on CMOS architecture and are manufactured using e-trimmed technology to minimise the errors associated with the input offset voltages, offset drifts and include EMI filters at their inputs. Table 1 compares some important parameters of OPA4192 at room temperature with other CMOS op-amps which are being used in cryogenic applications.

Table 1. Comparison of OPA4192 with other cryogenic CMOS op-amps used at ESO.

| Parameter | OPA4192 | TLC2274 | OPA4350 | OPA4354 |
|---|---|---|---|---|
| Gain bandwidth (MHz) | 10 | 2.2 | 38 | 100 |
| Voltage noise (nV/√Hz) | 5.5 | 9 | 5 | 6.5 |
| CMRR (dB) | 140 | 75 | 90 | 68 |
| Input bias current (pA) | ±5 | 1 | ±0.5 | ±3 |
| Slew rate (V/µs) | 20 | 3.6 | 22 | 150 |
| Quiescent current (mA) | 1.0 | 2.2 | 5.2 | 4.9 |
| Input offset voltage (µV) | ±10 | 300 | ±150 | ±2 |
| Offset drift (µV/°C) | ±0.2 | 2 | ±4 | ±4 |

In addition to its excellent DC precision and AC performances, the OPA4192 offers high capacitive load drive capability of up to 1nF, which makes it attractive for driving signals over long cable lengths for slow speed operation of the detectors.

## 3. WARM MUX TEST

Some of the performance advantages of the buffered mode could be readily verified with multiplexer at room temperature. Two boards have been prepared and tested: one board is configured for operating the detector in un-buffered mode and the other to run in buffered mode with individual external loads and supplies. The following sections show the performance differences between the two configurations. Please note when the preamp boards are swapped between the two configurations, the output mode register on the detector is programmed accordingly to select the correct output mode of operation.

**3.1 DC performance**

The DC transfer function of the signal chain from preamp to the ADC was measured by sweeping the external DC reference as one of the inputs of the differential preamp. The pixel source follower gain was measured by sweeping the VRESET voltage whist the detector is kept under reset. The output PMOS source follower gain was measured independently using the reference channel by sweeping DSUB voltage. Table 2 lists the DC gain, linearity of the preamp and the pixel source follower / output source follower gain of the detector as measured in un-buffered and buffered modes. The gain of the preamplifier and the pixel and output source followers are close to the designed and expected values.

Table 2. DC gain and non-linearity of the signal chain.

| Parameter | Unbuffered Mode | Buffered Mode |
|---|---|---|
| Preamp DC gain (non-linearity) | 7.76  (0.89%) | 7.77  (0.53%) |
| Output source follower gain | - | 0.896 (2.19%) |
| Pixel source follower gain | 0.971 (3.32%) | 0.997 (1.18%) |

## 3.2 Pixel frequency response

A correlated double frame with exposure to light revealed some defects along columns (Figure 3) and these were used for a quick assessment of pixel speed performance in both configurations under similar test conditions. The multiplexer is readout at 100kHz, 200kHz, 300kHz and 400kHz pixel rates.

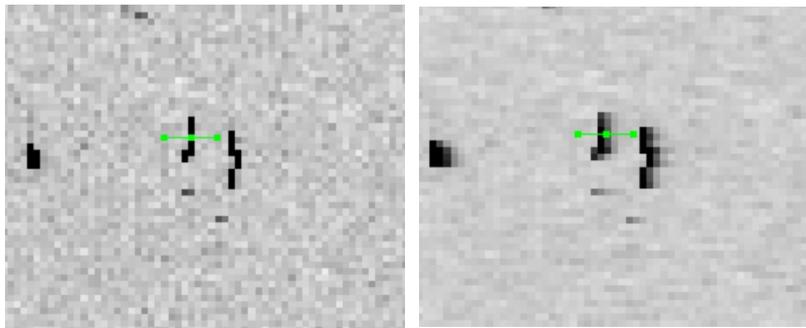

Figure 3. Defect pixels along columns used as a quick check for pixel speed performance. A plot across the defect column show the pixel speed performance. Left - buffered mode 200kHz, Right – unbuffered mode 200kHz.

In unbuffered mode, the pixel frequency response becomes poor and show charge smear to the neighboring columns in the readout direction beyond 100kHz pixel rate as seen in Figure 4, left panels. The smearing could be reduced by reducing the VBIASGATE which increases the current through the unit cell, but has a little effect beyond 200kHz. The faster the pixel rate, the more the charge smearing results. The pixels require longer settling times as they drive large combined capacitances including external capacitances to the preamplifier. In the case of buffered mode, no charge smear is noticeable up to 400kHz as shown in Figure 4, right panels. As expected, the on-chip output buffer isolates the dominant external capacitance that the pixel sees (in un-buffered case) and hence achieves higher pixel frequency.

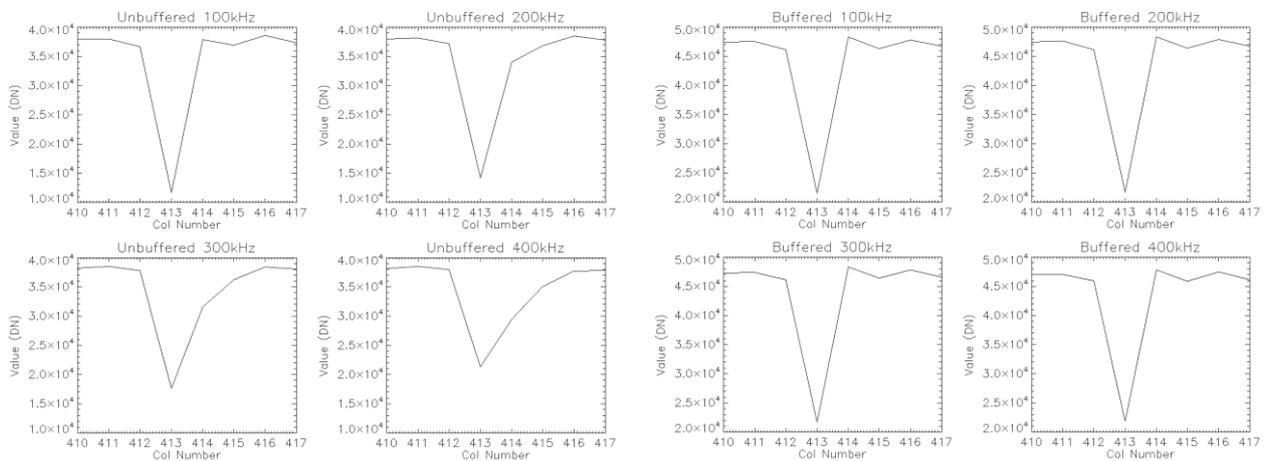

Figure 4. Plot across a defect pixel (along row) showing the performance difference between the un-buffered (left) and buffered (right) modes.

## 3.3 Electrical crosstalk

The noise in the normal readout channels is high at room temperature to measure any electrical crosstalk. However, from the readout electronics point of view, the reference output appears same as the normal outputs and hence it could be used to verify the electrical crosstalk. Crosstalk is measured using a window reset in that allows quick way of measuring crosstalk between outputs[2]. Flats are obtained with a signal of a few thousand electrons where a small windows region positioned in the middle of each readout channel is reset in a CDS frame. The flat with window reset is then subtracted from a similar flat without the window reset. The rows corresponding to the window region are binned to a single row in the resultant frame to improve the signal to noise ratio to verify the electrical coupling between readout channels.

Figure 5 left, below shows a step signal (in the windowed region) in one of the normal readout channels along with the signal in the reference channel when the detector is operated in un-buffered mode. A maximum of 0.2% crosstalk in the reference channel, corresponding to the edges of the step signal in a normal channel. The polarity of the coupling corresponding to the step function is dependent on the horizontal readout direction of the normal channel in relation to the reference channel.

The tests are repeated with the detector in buffered mode along with the individual external pullup supplies for each output on the preamp board. No noticeable crosstalk noticed (Figure 5, right) in this configuration at all speeds. The buffered mode operation of the detector with the corresponding external pull up supplies eliminates the electrical crosstalk. As the output impedance of the detector outputs in the buffered mode is low, no signal coupling between the outputs is noticed. See Section 5 for a modelling of crosstalk that shows the effect of the output source impedance in the signal coupling.

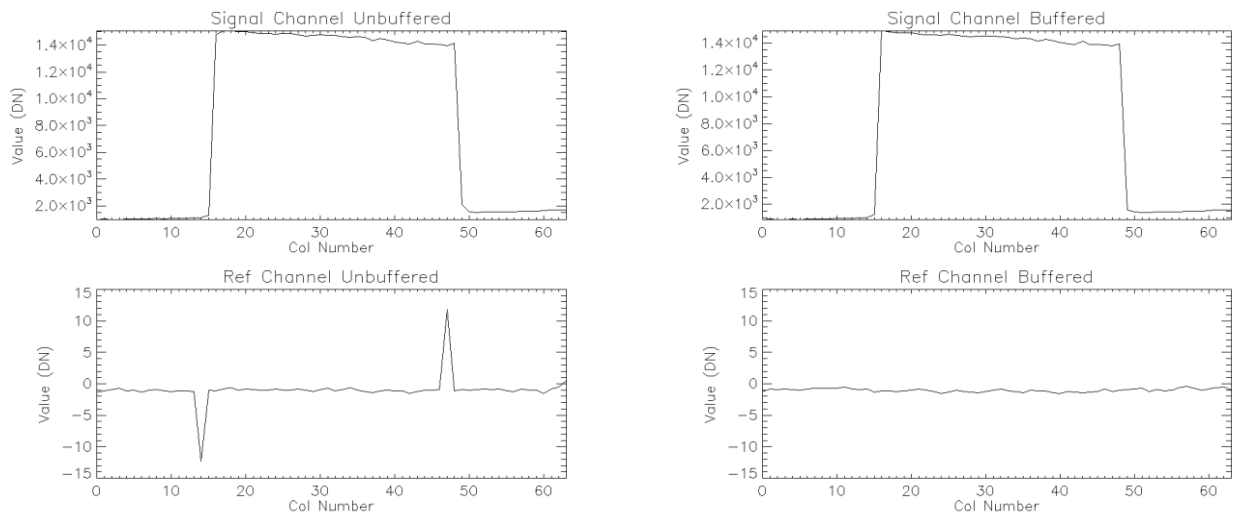

Figure 5. Electrical crosstalk between normal and reference outputs. Left – AC crosstalk at about 0.2% seen in un-buffered mode. Right – No crosstalk seen in the buffered mode.

## 3.4 Read noise

As the read noise in the normal channels is high at room temperature, the noise performance with pixel speed is measured using the reference channel in buffered mode. The read noise more or less remained same (about 5 to 6 counts) up to 400kHz pixel rates. The preamplifier board has an option to connect either the external DC voltage or the detector reference output as one of the inputs (offset) of the differential amplifiers of all channels. When the detector reference output is connected as the offset input, the differential amplifier corresponding to the reference channel sees the same signal (detector reference output) at its both inputs. As expected, the common mode noise at the preamplifier inputs is completely eliminated resulting noise data without any pickup or 1/f banding in the reference channel. The read noise in the reference channel is reduced to about 1.5 counts from about 5 – 6 counts with the external offset. The normal channels are too noisy at room temperature to see any difference in the noise performance in this mode.

# 4. COLD TESTS WITH SCIENCE H2RG DETECTOR

The preamplifier configured for a buffered mode of operation was prepared for vacuum and cryogenic use and mounted in its enclosure and integrated with a science detector in a test facility. The detector was operated at 80K and the preamp mount was held at about 60K for the measurements. The new op-amps (OPA4192) work well at 60K even from a cold start. An external offset is used as a reference input to the differential preamplifiers. The detector is operated with a bias of 0.25V. Photon transfer data obtained from just above bias level to signal saturation resulted a conversion gain of 2.19e/ADU and a full well of about 80ke. The preamp gain has been verified at cold to be same as at room temperature.

## 4.1 Pixel frequency response

Once again the window reset method is used to test the pixel speed response. In this case, an average frame with a reset window (33 columns x 257 rows) defined in a readout channel is subtracted from a reference flat without reset window. The rows of the window region are then averaged (in column direction) to improve signal to noise ratio. A plot along the averaged line shows transition corresponding to the window within on pixel. No signal lag (charge smear) is seen at the boundaries of the window region up to 500kHz pixel rate. The bias gate is kept at its nominal voltage of 2.4V.

The frequency response analysis showed that the first pixel at the start of the window region attained 99.3% of its final value at pixel frequencies up to 500kHz (Figure 6). The remining 0.7% is the charge diffusion at the boundaries due to the inter-pixel capacitance.

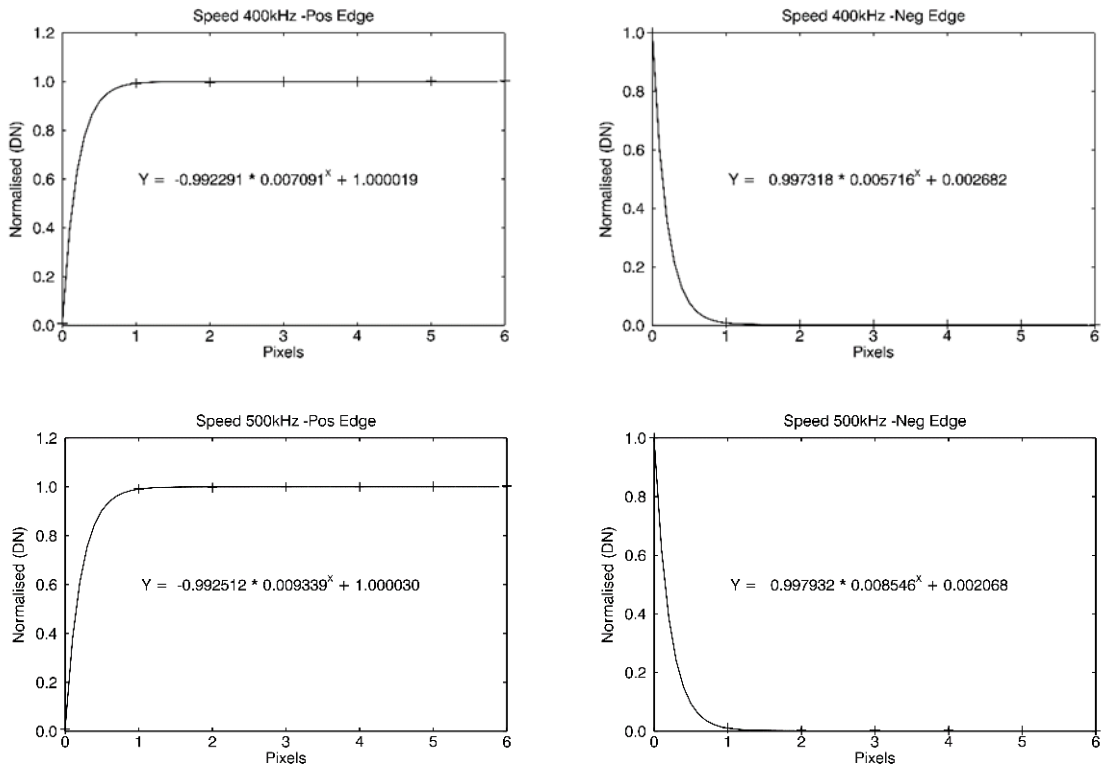

Figure 6. Pixel frequency response at 400kHz (top) and at 500kHz pixel rate (bottom).

## 4.2 Electrical crosstalk

Window reset method is used to measure the electrical crosstalk between different outputs. Figure 7 below shows no traces of electrical crosstalk in the output channel-3 when its neighbouring channel-2 is processing a high step signal. Similarly, no noticeable crosstalk is seen in other readout channels when high signal is processed in any other channel. A crosstalk matrix is generated by measuring crosstalk in all other channels for a window reset in each normal readout channel.

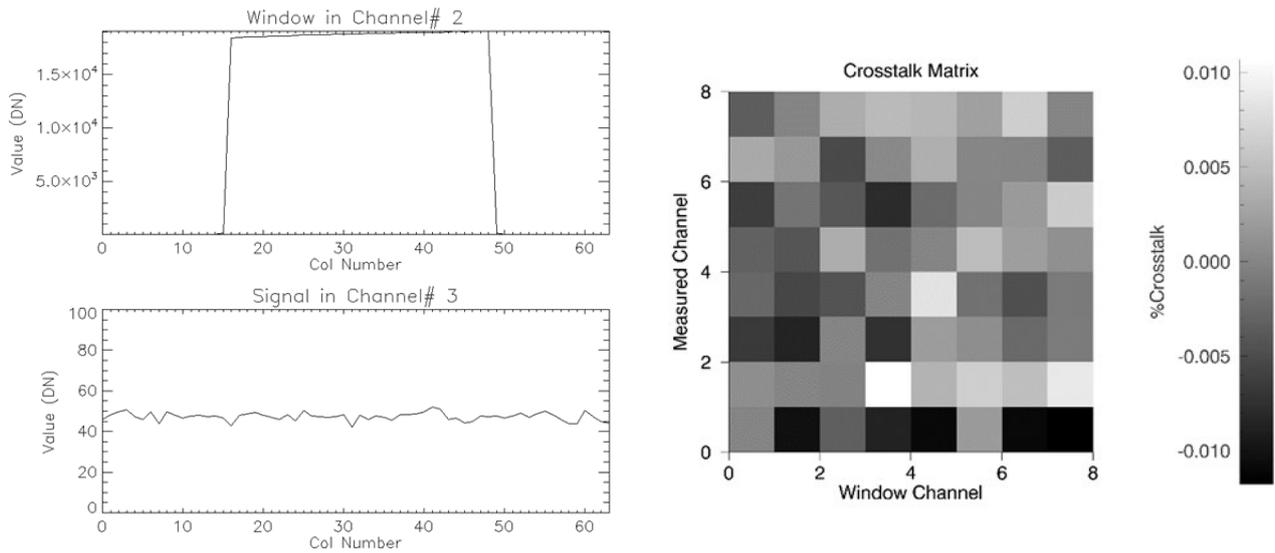

Figure 7. Electrical crosstalk between neighbouring channels and the matrix for all 8 signal channels.

### 4.3 CDS Read noise

The detector was kept under dark for a few hours before taking the data for read noise measurements. Two CDS bias frames are obtained at each pixel frequency. The two CDS frames are subtracted from each other and the resultant frame is divided by root(2). The read noise measured from a small region of 200x200 pixels. The CDS read noise remains almost same up to 500kHz pixel rates.

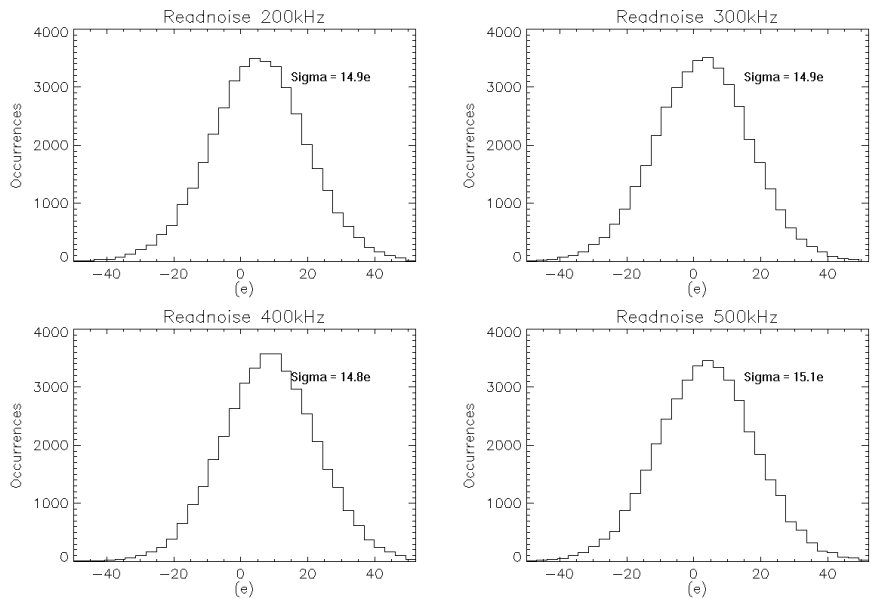

Figure 8. CDS read noise at different pixel rates.

The CDS read noise bitmap showed no horizontal banding or pick up noise up to 400kHz pixel rate. The bitmaps shown below is one quarter of the detector, corresponding to the readout channels 0 to 7. The bitmap display is scaled between -30e to +30e.

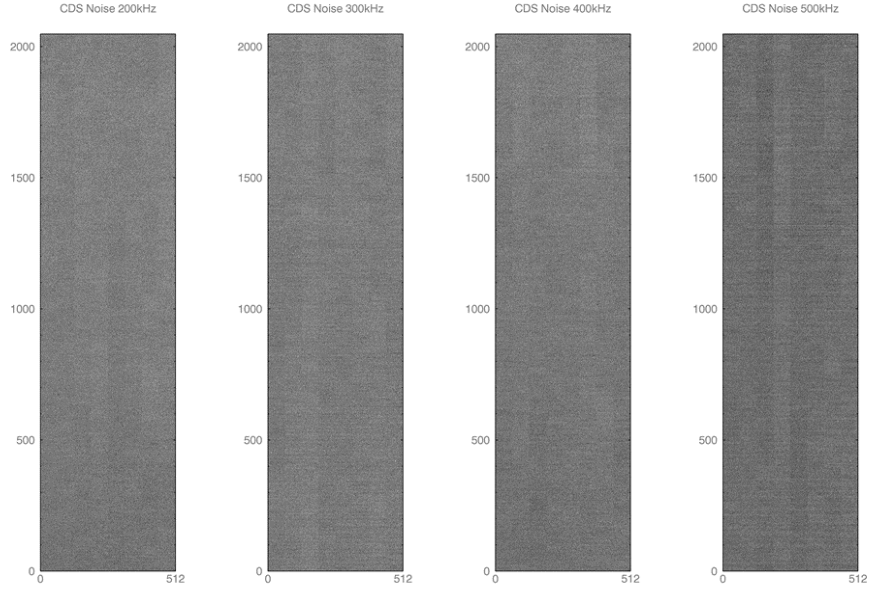

Figure 9. CDS read noise bitmaps at different pixel frequencies.

### 4.4 Fowler read noise

After keeping the detector in dark for several hours, two sets of NDR cube data of 1024 frames is obtained with DIT=1s. Only 256 rows of data obtained due to the file size limits. The two data cubes were then analysed for noise for different Fowler and NDR samples. Fowler pairs were created using subsequent frames in the cube. For example, Fowler-1 pair is created from the reads 1 and 2, Fowler-2 pairs created from the reads 1, 2 and 3, 4, Fowler-4 pairs is created from reads 1 to 4 and 5 to 8 etc. There is no gap between the two sets of reads of a Fowler pair. The read noise is obtained from two identical Fowler pairs created from the two sets of NDR cube data. Fowler noise is measured in a small region of the array. A noise minimum of 2.86e was achieved with 256 Fowler pairs. The noise starts to increase with further samples as the dark contributions start to dominate the noise.

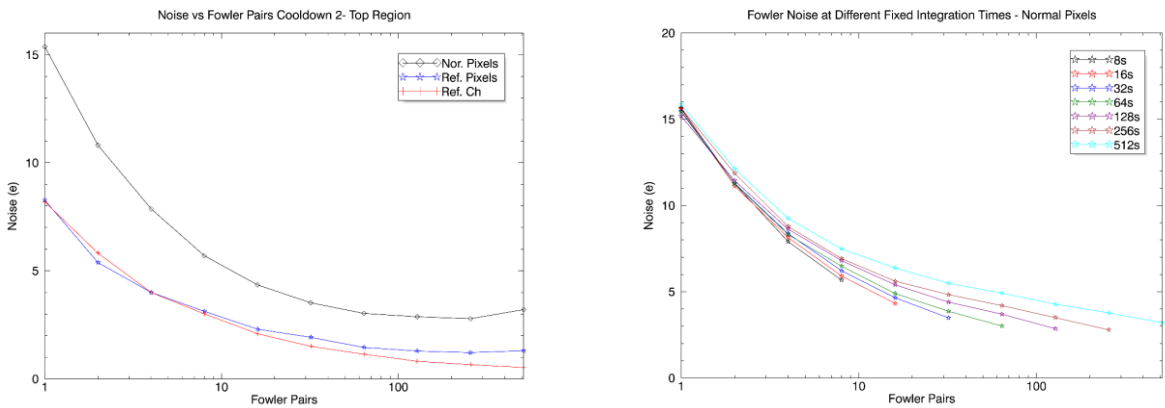

Figure 10. Left - Fowler Noise vs Fowler Pairs at 200kHz, Right - Fowler noise with fixed integration times.

From the same set of NDR data cubes, Fowler noise is also measured for different Fowler pairs separated by different fixed integration times. For example, for a 32s fixed integration time, Fowler-1 noise is measured using the read 1 and read 33, Fowler-2 noise is measured using reads 1,2 and 32 and 33, Fowler-4 noise is measured using reads 1 to 4 and reads 32 to 35 etc. Similarly for a 512s fixed integration time, Fowler-1 noise measured from read 1 and read 513,

Fowler-2 is measured from reads 1, 2 and reads 513, 514 etc. The total noise continues to come down with number of follower pairs for a fixed integration time.

## 4.5 Up the ramp read noise

The Up The Ramp (UTR) read noise computed from the same two read noise data cubes is shown in Figure 11. Once again the minimum UTR noise is obtained at 512 reads and then noise increases with further samples. Similar behaviour is seen on the embedded reference pixels, but the noise of the reference channel continues to decrease further. The minimum read noise is 5.14e at 512 reads. As expected, the minimum noise in the UTR sampling is approximately 1.8 worse than that of the Fowler read noise in the read noise regime.

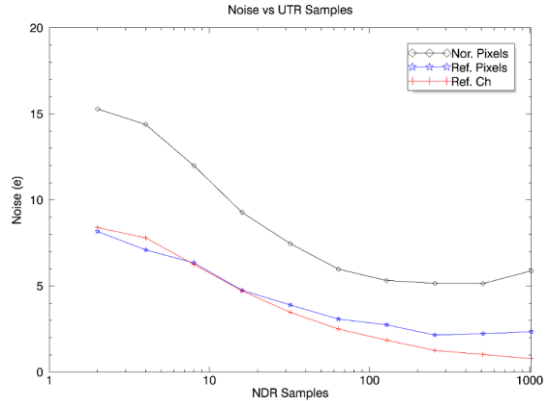

Figure 11. Up the ramp sampling noise vs number of samples.

## 5. ELECTRICAL CROSSTALK MODEL

In order to model the electrical crosstalk in the H2RG detector, two outputs are considered: one output driving high signal (source) and the other output which is affected due to the source (victim). The victim channel is held at a constant signal level.

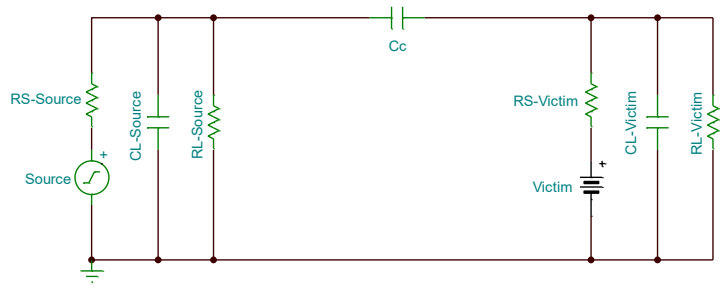

Figure 12. Source and victim model of the electrical crosstalk.

The output source impedance of the source channel (RS-Source) can be ignored as it is able to drive the signal up to the preamplifier and similarly the capacitive loads on the signal source (CL-Source) and on the victim output (CL-Victim) are ignored as the outputs are able to drive their respective capacitive loads. The source output load resistance (RL-Source) and the victim output load resistance (RL-Victim) are very high as they are connected to the non-inverting inputs of the differential preamplifier, and hence they can also be ignored. The coupling capacitance between the source and victim traces (Cc) is important as the signal coupling occurs via the electromagnetic filed due to the current flow in the source channel. Hence the above model can be simplified as shown in Figure 13.

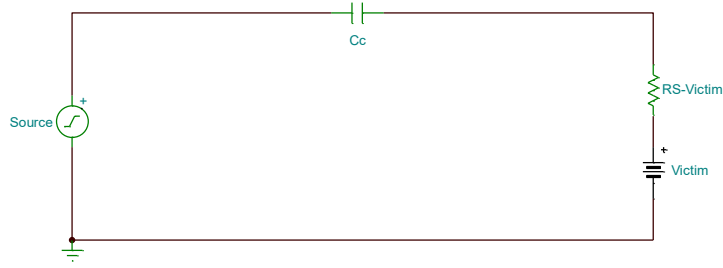

Figure 13. A simplified model for the electrical crosstalk between readout channels.

Since the crosstalk is capacitively coupled, the DC voltage of the victim can be considered as zero. So, the crosstalk is then just the ratio of RS-Victim to the total circuit impedance (RS-Victim + Reactance of Cc).

$$\text{Crosstalk} = \frac{\text{RSVictim}}{\text{RSVictim} + \frac{1}{j\omega C}}$$

In the case of un-buffered mode of operation, the impedance of the source / victim outputs is expected to be about 5k ohms at 70K. The amplitude of the signal coupling to the victim then depends on the rise / fall time of the source signal, the coupling capacitance, victim source impedance and the amplitude of the source signal. The reactance of the coupling capacitance is estimated at the maximum frequency corresponding to the rise and fall time of the source signal (f = 1/(2*Tr)). In the case of the buffered outputs, then the output source impedance is of the order of 150 ohms at 70K and hence the crosstalk is very low or negligible.

## 6. CONCLUSIONS

The goals of the test preamplifier design have been successfully tested with a H2RG detector which confirmed the advantages of operating the HxRG detectors in slow buffered mode compared to the conventional un-buffered mode. The buffered mode allows reading the detector at higher pixel rates with almost a flat noise performance whilst also eliminating the electrical crosstalk between the readout channels. A new, low power, low temperature co-efficient op-amps have been successfully tested at cryogenic temperatures which showed a superior performance compared to the currently used op-amps for slow speeds. In the context of the ELT instruments, the buffered mode operation allows to achieve a better performance whilst requiring only half the detector cryogenic electronics and detector warm controller electronics, which significantly reduces the associated cryo-mechanical complexities of the instruments. Similarly, based on this work, a new preamplifier has been designed for MOONS H4RG near infrared detectors and the results are being presented at this conference[3].